\documentclass[letter,draftclsnofoot,onecolumn,12pt,oneside]{IEEEtran}
\usepackage{cite}
\usepackage{epsfig,graphics,mathrsfs,amsmath,bm,color,yfonts,txfonts,slashbox,multirow}
\usepackage{subfigure}
\usepackage{algorithm2e}

\begin{document}

\title{Resource Allocation Optimization for Users with Different Levels of Service in Multicarrier Systems}
\author{Mirza Golam~Kibria and Shan~Lin
  \thanks{Copyright (c) 2015 IEEE. Personal use of this material is permitted. However, permission to use this material for any other purposes must be obtained from the IEEE by sending a request to pubs-permissions@ieee.org.}
 \thanks{The authors are with Wireless Network Research Institute, National Institute of Information and Communications Technology (NICT), Yokosuka Research Park, Japan 239-0847 (e-mails: $\left\{\rm{mirza.kibria,shanlin}\right\}$@nict.go.jp). }
 \thanks{This work has been completed when M.~G.~Kibria was with Graduate School of Informatics, Kyoto University, Japan 606-8501.}
 }
\maketitle

\begin{abstract}
We optimize the throughput of a single cell multiuser orthogonal frequency division multiplexing system with proportional data rate fairness among the users. The concept is to support mobile users with different levels of service. The optimization problem is a mixed integer nonlinear programming problem, which is computationally very expensive. We propose a novel and efficient near-optimal solution adopting a two-phase optimization approach that separates the subcarrier and power allocation. In the first phase, we relax the strict proportional data rate requirements and employ an iterative subcarrier allocation approach that coarsely satisfies desired data rate proportionality constraints. In the second phase, we reallocate the power among the users in an iterative way to further enhance the adherence to the desired proportions by exploiting the normalized proportionality deviation measure. The simulation results show that the proposed solution exhibits very strong adherence to the desired proportional data rate fairness while achieving higher system throughput compared to the other existing solutions. 
\end{abstract}

\IEEEpeerreviewmaketitle
\vspace{-2mm}
\section{Introduction}
\label{section:chapter3-1}
\IEEEPARstart{O}{rthogonal} frequency division multiple access (OFDMA) is a promising technique for achieving high downlink capacity \cite{Lawrey,Sternad}. The problem of allocating subcarriers and power to the users in OFDMA system has been an area of active research. In non-adaptive and fixed resource allocation schemes such as single user orthogonal frequency division multiplexing (OFDM) system and/or under static subcarrier allocation schemes such as frequency division multiple access (FDMA) and time division multiple access (TDMA), each user is allocated with an independent dimension without considering the channel state information (CSI). As a result, in such systems, the optimization problem reduces to only bit loading or power allocation. Unlike single user OFDM and static FDMA/TDMA, subcarrier allocation itself plays an important role in OFDMA systems in optimizing the system throughput due to multiuser diversity. This is due to the fact that the fading conditions for different users in the system are mutually independent, and the probability of a subcarrier being in the deep fade for all the users is very low \cite{Sternad}. Therefore, efficient radio resource management schemes are very important to maximize the OFDMA system throughput.

There have been many works in literature dealing with the problem of resource allocation in OFDMA system under various system constraints \cite{Wong,Rhee,Jang,Yu, Ng,Hu, Xia}. In \cite{Wong}, the authors proposed an iterative resource allocation algorithm to minimize the total transmitting power under fixed user data rates and bit error rate constraints. The algorithm proposed in \cite{Jang} is aimed at the maximization of data rate under total transmitting power and target bit error rate requirements. In \cite{Yu,Ng}, the authors claim that nonconvexity is not an issue for the resource allocation problem in OFDMA system if the number of subcarriers is very large. The authors of \cite{Jang,Rhee} have shown that the overall system capacity of OFDMA system is optimized when each subcarrier is assigned to the user with the best channel gain. In \cite{Hu,Xia}, adaptive resource allocation in OFDMA system is considered under partial CSI.

Fairness among the users is a very important issue, which ensures that all the users are able to achieve their required data rates, as in a system with quality of service requirements. In most of the practical scenarios, different users have different data rate requirements with different types of services and fees. 
There is a special category of rate adaptive and dynamic resource allocation approaches \cite{Shen,Wong1,Sadr1,Mohanram}, which support variable data rate services with fairness in the system and balances the trade-off between capacity and fairness. In this category, the objective is to maximize the system throughput under total transmitting power constraint, and the target is to maintain proportionality between users' achieved data rates rather than satisfying specific requested data rates. 

 In \cite{Shen}, the authors proposed a solution (referred to as the ROOT-FIND algorithm) that ensures the rates of different users are proportional. Though the solution exhibits very good adherence to the proportional data rate fairness requirements, the algorithm needs to solve a set of nonlinear equations, which are computationally very expensive. Furthermore, this solution is based on a high signal-to-noise ratio (SNR) assumption, and as a result, does not perform satisfactorily in the low-SNR region. In \cite{Wong1,Sadr1}, the authors proposed suboptimal solutions (former one referred to as the LINEAR solution), which relax the desired user rate proportionality constraints and achieves acceptable proportional data rate fairness. A method for joint subcarrier and power allocation with proportional data rate fairness algorithm (referred to as the JSPA-WF algorithm) is proposed in \cite{Mohanram}. Though it provides a better throughput, like\cite{Wong1}, the proportional data rate fairness measures are also relaxed, i.e., does not exhibit strong adherence as desired.

In this letter, we propose a novel near-optimal solution for maximizing the throughput while strongly maintaining the proportional data rate fairness among the users. A disjoint subcarrier and power allocation approach in invoked. Unlike \cite{Shen}, the proposed solution does not require to solve a set of nonlinear equations, and performs equally well both in the high-SNR and in the low-SNR region. The proposed solution provides the overall system throughput as high as that in \cite{Mohanram}, and does not suffer much from the capacity loss.

\section{System Model and Problem Statement}
\label{section:chapter3-2}
Let us consider an OFDMA downlink system with a base station supporting data traffic to $K$ non-cooperative users with a single receiving antenna each. 
The data transmissions to different users are assumed to be subject to slowly-varying, independent frequency-selective Rayleigh fading. Perfect CSI is assumed to be available at the base station and a non-sharing subcarrier allocation scheme is considered, i.e., a subcarrier can be allocated to a single user only. Note that the assumptions of perfect CSI and no intra/inter-cell interference would generally not hold in practice. However, we relax these assumptions in this work for the sake of simplified performance analysis.
The data transmission is subject to regulated total transmitting power constraint, $P_{\rm{total}}$. The capacity achieved by user $k$ when transmitting data over subcarrier $n$ is given by
\begin{equation}
\label{optzm}
r_{kn}=N^{-1}{\rm{log_2}}\left( 1+{{{p}_{kn}}|z_{kn}|^2}/\left(N_0B/N\right) \right)\hspace{5mm}\text{in bps/Hz},
\end{equation}
where $N$ is the number of subcarriers, $B$ is the system bandwidth, and  $z_{kn}$ defines the frequency gain on subcarrier $n$ of user $k$. $N_0 B/N$ is the variance of additive white Gaussian noise (AWGN) over subcarrier $n$, where $N_0$ is the noise power spectral density. The quantity $|z_{kn}|^2/\left(N_0 \frac{B}{N}\right)$ is defined as the SNR of subcarrier $n$ if it is allocated to user $k$. $p_{kn}$ is the power allocated to subcarrier $n$ of user $k$. Lets denote $c_{kn}$ the association variable, i.e. $c_{kn}=1$ if subcarrier $n$ is assigned to user $k$, 0 otherwise. The total rate achieved by user $k$ is given by $R_k=\sum\limits_{n=1}^{N}{{{{c}}_{kn}}{{r}_{kn}}}$ in bps/Hz. 

The different levels of service among the users can be embedded by introducing a set of nonlinear constraints, $R_1:R_2:\cdots:R_K=\varphi_1:\varphi_2:\cdots:\varphi_K$, where $\{\varphi_1,\varphi_2,\cdots,\varphi_K\}$ are the parameters that enforce proportional fairness or desired levels of service among the users. We can ensure arbitrary proportional throughput or levels of service among the users by varying the values of these predetermined parameters. The advantageous fact about embedding these proportional data rate fairness measures is that we can explicitly control the data throughput ratio among the users. All the calculations relating to resource allocation are performed at the base station.

In practical, the regulatory scenario enforces a total transmitting or radiated power constraint. The main objective of our optimization problem is to perform optimal resource allocation in order to achieve the maximum system throughput under total transmitting power constraint and proportional data rate fairness among the users. The optimization problem is cast as
\begin{equation}
\label{main}
\begin{array}{*{35}{l}}
\underset{{{c}_{kn}},{{p}_{kn}}}{\mathop{\max }}\,\sum\limits_{k=1}^{K}R_k \vspace{1.5mm} \\
\vspace{2mm}
\text{}\text{subject to} \hspace{.70mm}\text{~C1:}\sum_{k=1}^{K}c_{kn}\le 1,\hspace{1mm}{{c}_{kn}}\in \left\{ 0,1 \right\},\left( \forall k,n \right); \vspace{1.5mm} \\
\text{}\hspace{15mm} \text{C2:}{{p}_{kn}}\ge 0,\left( \forall k,n \right); \vspace{2mm} \\
\text{}\hspace{15mm} \text{C3:}\sum\limits_{k=1}^{K}{\sum\limits_{n=1}^{N}{{{c}_{kn}}{{p}_{kn}}\le {{P}_{\rm{total}}}}}; \vspace{1.5mm} \\
\text{}\hspace{15mm} \text{C4: }\frac{R_1}{{{\varphi }_{1}}} =\frac{R_2}{{{\varphi }_{2}}}=\cdots =\frac{R_K}{{{\varphi }_{K}}}. \vspace{1.5mm} \\
\end{array}
\end{equation}
The optimization problem in \eqref{main} is a mixed integer nonlinear programming (MINLP) problem with $KN$ binary integer variables. MINLP problems are the most general class within algebraic optimization. The computational complexity of the problem in \eqref{main} grows exponentially with the number of integer variables. MINLP problems are undoubtedly very difficult to solve as they accumulate all the intrinsic complexities of both of their subclasses: the complexity in solving nonconvex NLP and the combinatorial nature of mixed integer optimization programs. In the next section, we propose a near-optimal and low-complexity (compared to the original MINLP problem) solution for the aforementioned optimization problem.

\section{Proposed Near-optimal Solution}
\label{section:chapter3-3}
The resource allocation process is divided into two phases, (i) subcarrier allocation phase and (ii) power reallocation phase. The suboptimal subcarrier allocation maintains the coarse proportionality of user data rates with equal power allocation over all the subcarriers. A non-overlapping subcarrier allocation scheme is employed, i.e., a subcarrier sharing is not allowed. 

\vspace{-2mm}
\subsection{Subcarrier allocation phase}
Let $\mathcal{S}_k$ be the set of subcarriers allocated to user $k$ with $\mathcal{S}_k \cap \mathcal{S}_l=\emptyset; \hspace{1mm}k,l=1,\cdots,K \hspace{1mm}(k\ne l)$ while $N_k$ being its cardinality, i.e., $N_k={\rm{cardinality}}\hspace{1mm}(\mathcal{S}_k)$. Therefore, the corresponding power allocated to user $k$ after this subcarrier allocation is given by $N_kP_{\rm{total}}/N$. In particular, we follow and employ the subcarrier allocation strategy presented in $\bf{Algorithm}~\ref{algo2}$, which is the modification of the subcarrier allocation scheme discussed in \cite{Rhee} so as to incorporate proportional data rate fairness measures. Let $x_{ij}$ defines the $(i,j)$th element of matrix $\bm{X}$ while ${x}_{i}$ defines the $i$th element in vector $\bm{x}$, and $\mathbb{R}$ defines a real space with appropriate dimension.
 \begin{algorithm}
  \SetAlgoLined
 \textbf{Initialization}: $\mathcal{N}=\left\{ 1,2,3,\cdots ,N \right\}$,  $\mathcal{S}_k=\emptyset$, $p={P}_{\rm{total}}/N$;\
  $\bm{H}=[h_{kn}]_{K\times N}\text{ with }{h}_{kn}=|z_{kn}|^2/\left(N_0 \frac{B}{N}\right)$; \
  
  \BlankLine
  \For{$k=1{\rm{~to~}} K$}{
    $\text{Let~}h={\mathop{\max }}\,\{{h}_{kn},n\in \mathcal{N}\}$; $\text{Update~}{{S}_{k}}={{S}_{k}}\cup \left\{ n \right\}$\;
        $\text{Calculate~}{{R}_{k}}={{\log }_{2}}(1+ph)$; $\text{Update~}\mathcal{N}=\mathcal{N}-\left\{ n \right\}$\;
       }
  \While{$\mathcal{N}\ne \emptyset$}{
   $\text{Find~}i\text{~such that~} i = \arg \underset{k\in\{1,\cdots,K\}} {\mathrm{min}} ~R_k/\varphi_k$\;\vspace{1mm}
     $\text{For~}i\text{,~find~}n\text{~such that~}n = \arg \underset{m\in\mathcal{N}} {\mathrm{max}} ~h_{im}$\;\vspace{1mm}
       $\text{Update~}{\mathcal{S}_i}={\mathcal{S}_i}\cup \left\{ n \right\};{\rm{~}}\mathcal{N}=\mathcal{N}-\left\{ n \right\}$\;\vspace{1mm}
              $\text{Calculate~}{{R}_{i}}=\sum\nolimits_{n\in {\mathcal{S}_i}}{{{\log }_{2}}(1+p{h}_{in})}$\;\vspace{2mm}
                              }
                                \BlankLine
      \caption{Suboptimal subcarrier allocation}
   \label{algo2}
\end{algorithm}
The fundamental feature of this suboptimal subcarrier allocation algorithm is that each user uses the subcarriers with high SNR as much as possible. At each iteration, user $k$ with the lowest or minimum normalized capacity, i.e., $R_k/\varphi_k \le R_i/\varphi_i, \forall i, 1\le i \le K$ has the opportunity to select the subcarrier that has the maximum gain for it. Since equal power allocation over all the subcarriers is considered, the subcarrier allocation algorithm is suboptimal and achieves coarse proportional data rate fairness, i.e., ${{R}_{i}}:{{R}_{j}}\approx{{\varphi}_{i}}:{{\varphi}_{j}},\hspace{3mm}\forall i,j\in \left\{ 1,\cdots ,K \right\},i\ne j$. 

It might seem from the {\bf{for loop}} that {\bf{Algorithm 1}} will yield to different results if the chosen order of the users is changed. However, we have noticed that the order of the users has almost no impact on the achieved proportional data rates of the individual users, and as a result, on the overall system throughput. The reason is that the likelihood of any particular subcarrier exhibiting the highest gain or deepest fading simultaneously for all the users is very low because of multiuser diversity prevailed by variations across the frequency, time and spatial domains of the user channels in an OFDMA system. No matter which order of the users we choose, will eventually lead to same subcarrier allocation. 
 
\subsection{Power reallocation phase}
\label{section:chapter3-5}
In this section, we discuss the power reallocation phase that is employed to enhance the adherence to the desired proportional data rate fairness measures. The proportionality deviation, i.e., the difference between the desired proportion and the achieved proportion is given by 
\begin{equation}
\label{error}
{\bm{\xi}}=\left\{\left( \frac{{{\varphi }_{1}}}{{\Sigma}_{\varphi}}-\frac{{{R}_{1}}}{{\Sigma}_{R}} \right),\left( \frac{{{\varphi }_{2}}}{{\Sigma}_{\varphi}}-\frac{{{R}_{2}}}{{\Sigma}_{R}} \right),\cdots,\left( \frac{{{\varphi }_{K}}}{{\Sigma}_{\varphi}}-\frac{{{R}_{K}}}{{\Sigma}_{R}} \right)\right\},
\end{equation}
where ${{\xi}}_k=\left( \frac{{{\varphi }_{k}}}{{\Sigma}_{\varphi}}-\frac{{{R}_{k}}}{{\Sigma}_{R}} \right)$ is the normalized proportionality deviation of user $k$, ${\Sigma}_{\varphi}={\sum\limits_{k=1}^{K}{{{\varphi }_{k}}}}$ and ${\Sigma}_{R}={\sum\limits_{k=1}^{K}{{{R }_{k}}}}$. The sign of ${{\xi}}_k$ defines whether user $k$ has been allocated a lower or a higher proportional capacity than the desired proportions. The basic idea of this iterative power reallocation is that we reduce the power of the user that has achieved a higher proportional rate than the desired value, and add this power to the user that has a lower proportional rate than the desired value. The parameter ${{\Delta }}=\frac{1}{K}\sum\limits_{k=1}^{K}{| {{\xi}}_k |}$ has been used to quantify the average normalized proportionality deviation measure. Our objective is to make ${{\Delta }}$ as small as possible. The heuristic routine employed in the power reallocation process in given in {\bf{Algorithm 2}}.
\begin{algorithm}[h]
  \SetAlgoLined
  \textbf{Initialization}: $\bm{H}= [h_{kn}]_{K\times N}\text{ with }{h}_{kn}=|z_{kn}|^2/\left(N_0 \frac{B}{N}\right)$ \
  $\bm{P}= [p_{kn}]_{K\times N},\hspace{1mm} \delta=P_{\rm{total}}/8N ,\hspace{1mm}N_{\rm{iter,max}},\hspace{1mm}N_{\rm{iter}}=0; $
  \BlankLine
  \While{$N_{\rm{iter,max}}>0$}{
   $ \text{Find~}r=\arg \underset{k\in\{1,\cdots,K\}} {\mathrm{max}} ~\left({{\xi}}_k \right); \hspace{1mm}s=\arg \underset{k\in\{1,\cdots,K\}} {\mathrm{min}} ~\left( {{\xi}}_k \right)$\;\vspace{1mm}
     $ \text{Calculate~}{{R}_{i}}=\frac{1}{N}\hspace{-1mm}\sum\limits_{n\in {\mathcal{S}_i}}{{{\log }_{2}}(1+{{{p}}_{in}}{{{h}}_{in}})},i\in\{r,s\}$\;\vspace{1mm}
      $ \text{Calculate~}{{\Delta }_{\rm{old}}}=\frac{1}{K}\sum\limits_{k=1}^{K}{| {{\xi}}_k |},\hspace{1mm} R_x=R_r,\hspace{1mm}R_y=R_s$\;\vspace{1mm}
       $ \text{Set~}{{\bm{p}}_{s}}=\left\{ {{{p}}_{sz}}\hspace{1mm}(\forall z,z\in {\mathcal{S}_s}) \right\};\hspace{1mm}{{\bm{p}}_{r}}=\left\{ {{{p}}_{rz}}\hspace{1mm}(\forall z,z\in {\mathcal{S}_r}) \right\}$\;\vspace{2mm}
        $ \text{Calculate~}{{\Sigma }_{s}}=\sum\limits_{n\in {{S}_{s}}}{{{{p}}_{sn}}+\delta };\hspace{1mm}{{\Sigma }_{r}}=\sum\limits_{n\in {\mathcal{S}_r}}{{{{p}}_{rn}}-\delta }$\;\vspace{1mm}
         $ \text{Calculate~}{{\lambda }_{i}}=\left( {{\Sigma }_{i}}+\sum\limits_{n\in {\mathcal{S}_i}}{\frac{1}{{{{h}}_{in}}}} \right)/{{N}_{i}},\hspace{1mm}i\in \{r,s\}$\;\vspace{1mm}
          $ \text{Set~} \bm{p}_{i}^{\rm{new}}=\left\{ {{\omega }_{1}},{{\omega }_{2}},\cdots ,{{\omega }_{{{N}_{i}}}} \right\} ; \hspace{1mm}{{\omega }_{j}}={{\left( {{\lambda }_{i}}-\frac{1}{{{{h}}_{i,\mathcal{S}_i^{(j)}}}} \right)}^{+}}$\;
           $ \text{~~~with~~}\sum\limits_{j=1}^{N_i}{{\omega }_{j}}= {{\Sigma }_{i}},\hspace{1mm}f(x)={{(x)}^{+}}:=f(x)=\left\{ \begin{matrix}
   0,\hspace{1mm}x<0  \\
   x,\hspace{1mm}x\ge 0  \\
\end{matrix} \right.$\
            $ \text{~~~Here,~~}\mathcal{S}_k^{(i)}\text{~denotes the~}i\text{th~element in~}\mathcal{S}_k$\;\vspace{2mm}
             $ \text{Set~}{{{p}}_{iz}}\hspace{1mm}(\forall z,z\in {\mathcal{S}_i})=\bm{p}_{i}^{\rm{new}},\hspace{1mm}i\in \{r,s\}$\;\vspace{2mm}
              $\text{Calculate~} {{R}_{i}}=\frac{1}{N}\sum\limits_{n\in {\mathcal{S}_i}}{{{\log }_{2}}(1+{{{p}}_{in}}{{{h}}_{in}})},\hspace{1mm}i\in\{r,s\}$\;\vspace{1mm}
               $ {\Sigma}_{R}={\sum\limits_{k=1}^{K}{{{R }_{k}}}};\text{ Calculate }{\bm{\xi}} \text{ using }\eqref{error};\hspace{1mm}{{\Delta }_{\rm{new}}}=\frac{1}{K}\sum\limits_{k=1}^{K}{| {{\xi}}_k |}$\;\vspace{2mm}
                  \eIf{${{\Delta }_{\rm{new}}}\ge{{\Delta }_{\rm{old}}}$}{
      $R_r=R_x, \text{~}R_s=R_y; {{{P}}_{i,z}}\hspace{1mm}(\forall z,z\in {\mathcal{S}_i})=\bm{p}_{i},\hspace{1mm}i\in \{r,s\}$\;
      ${\textbf{break}}$\;
      }{
      $N_{\rm{iter,max}}=N_{\rm{iter,max}}-1$; $N_{\rm{iter}}=N_{\rm{iter}}+1$\;
      }
    }
  \caption{Power reallocation process}
   \label{algo}
\end{algorithm}

The operational mechanism of the proposed power reallocation process is as follows. The power reallocation process enhances the adherence to the desired data rate proportions in an iterative way. In every iteration, the power of the two users with the most unfair proportions is exchanged. The values in ${\bm{\xi}}$ define the normalized deviation of the proportion of all the users from the desired proportions. The overall proportional data rate fairness of the whole system will be enhanced when the value of ${{\Delta }}=\frac{1}{K}\sum\limits_{k=1}^{K}{| {{\xi}}_k |}$ becomes smaller. In the ultimate proportional data rate fairness scenario, the value of $\Delta$ is 0. 

First, the two users with the most unfair normalized proportionality deviations are chosen. The indices $s$ and $r$ denote the users with the most unfair normalized proportionality deviations, in terms of achieving the lowest and the highest normalized throughputs, respectively, when compared to the desired normalized throughput proportions. Therefore, user $s$ needs additional power to enhance its achieved data rate and to match its desired proportion while user $r$ needs to release some extra power to bring the achieved additional data rate down. To perform this power exchange operation smoothly, we add and subtract a very small amount of power, which is quantified as $\delta$ to users $s$ and $r$, respectively. Thereafter, the new power, ${{\Sigma }_{s}}$ and ${{\Sigma }_{r}}$ are distributed optimally among the allocated subcarriers of users $s$ and $r$, respectively, to optimize their individual capacities employing the standard water-filling algorithm. In order to ensure the convergence, the iterative process continues as long as there is an improvement in the proportional fairness measures.

 \section{Discussion and Performance Analysis}
 \label{section:chapter3-7}
 We consider an OFDMA system with a transmitter supporting data transmission to $10$ non-cooperating users. An OFDM scheme with $N=64$ subcarriers is employed. The power delay profile of the channel model we consider is an one-sided exponential profile, of which the relative powers of 6 delay taps are provided in Table \ref{Simulate} along with the other simulation parameters employed in this work. 
  
  \begin{table}[h]
\centering
\caption{Simulation Parameters} 
\label{Simulate}
\begin{tabular}{l |r}
\hline
\hline
\hspace{-2mm}Power constraint, ${P}_{\rm{total}}$ & \hspace{-2mm}1-5 W\\ 
\hspace{-2mm}AWGN spectral density, $N_0$ &\hspace{-2mm} -170 dBm/Hz\\
\hspace{-2mm}Number of subcarriers, $N$ & \hspace{-2mm}64\\
\hspace{-2mm}Bandwidth, $B$ & \hspace{-2mm}1 MHz\\ 
\hspace{-2mm}Channel Model & \hspace{-2mm}6-taps frequency-selective\\ 
\hspace{-2mm}Power delay profile & \hspace{-2mm}Exponentially decaying\\
\hspace{-2mm}Relative power of 6-taps & \hspace{-1.95mm}[0, -4.35, -8.69, -13.08, -17.43, -21.78] dB\\ 
\hline
\end{tabular}
\end{table}
 
 \begin{figure}[h!]
  \centering
   \includegraphics[scale=.51]{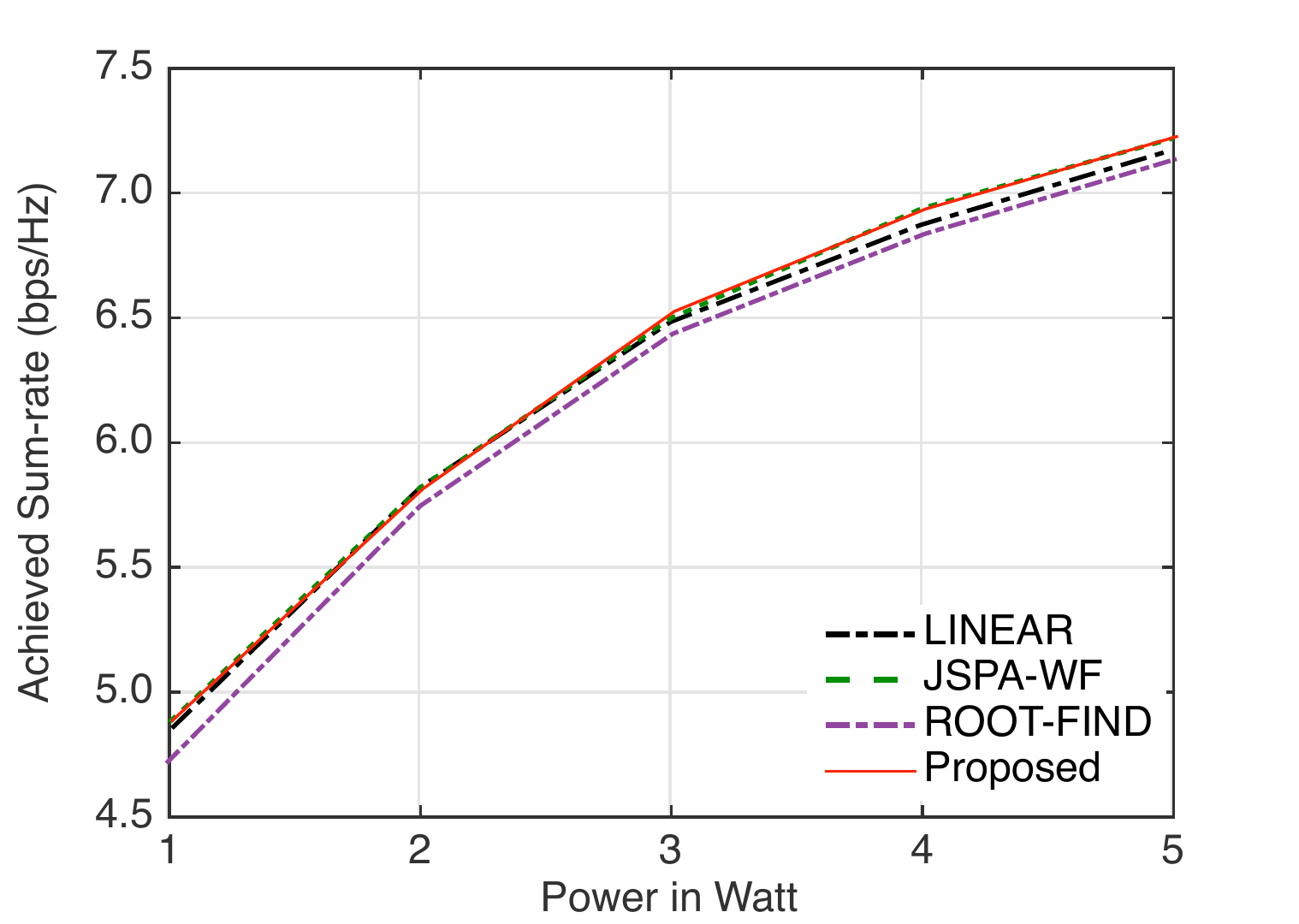}
   \caption{Sum-rates comparison when transmitting power, $P_{\rm{total}}$ is varied.   }
   \label{power user}
 \end{figure}
In Fig.~\ref{power user}, we evaluate the performance of our proposed solution when the transmitting power limit is swept from 1 Watt to 5 Watts, and compare it with other existing solutions. The ROOT-FIND solution suffers in the low-SNR region as the algorithm is formulated based on a high-SNR approximations. When the transmitting power is high, the ROOT-FIND algorithm performs better than the LINEAR solution. Whereas, our proposed solution perform equally well for both low-SNR and high-SNR scenarios.

 \begin{figure}[h!]
  \centering
   \includegraphics[scale=1.14]{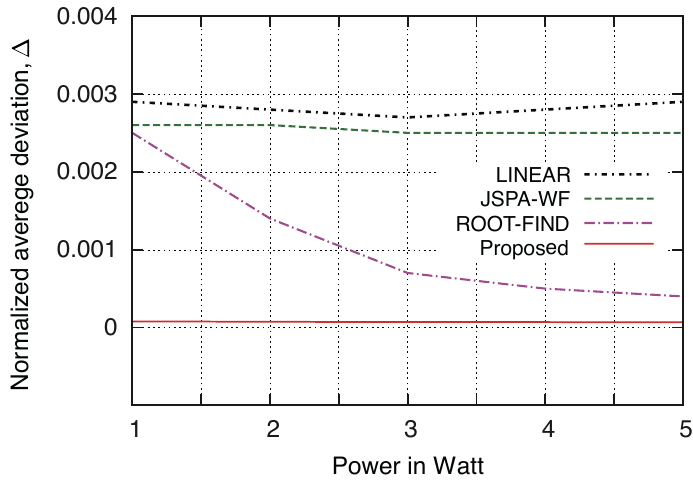}
   \caption{Behavior of average normalized proportionality deviation measure, $\Delta$ when maximum  transmitting power, $P_{\rm{total}}$ is varied. Here, $K=$10 and $\delta=P_{\rm{total}}/8N$.  }
   \label{Power Distortion}
 \end{figure}
 In Fig.~\ref{Power Distortion}, we evaluate the influence of varying the transmitting power on the normalized proportionality deviation measure $\Delta$ for the investigated algorithms. The LINEAR and JSPA-WF solutions show almost flat responses over all the transmitting power levels, but exhibit the worst adherences when compared to other existing solutions. Note that the ROOT-FIND solution suffers in the low-SNR region since the algorithm is formulated based on a high-SNR approximations. However, when the transmitting power is high, the ROOT-FIND algorithm performs better than the LINEAR and JSPA-WF solutions. The proposed solution has also almost consistent behavior of the parameter $\Delta$ over varying $P_{\rm{total}}$, but shows the best adherence to the desired data rate proportions. Unlike the ROOT-FIND solution, the proposed solution works equally well over the low-SNR and high-SNR regions. Note that  $\delta$ has noticeable impact on the iterative power reallocation phase performance. If a higher value of $\delta$ is chosen, the process requires less number of iterations to reach the steady state value of $\Delta$. Smaller $\delta$ will yield to large number of iterations while giving lower $\Delta$, i.e., better proportional fairness. Therefore, lower the value of $\delta$, better the adherence to the desired levels of service. However, decreasing $\delta$ below some threshold, will have no significance impact.

The computational complexity of our proposed algorithm is composed of two parts, namely (i) subcarrier allocation in {\bf{Algorithm 1}} with complexity of $\mathcal{O}(KN)$, and (ii) standard water-filling needs to be performed $2N_{\rm{iter}}$ times in {\bf{Algorithm 2}}. On the other hand, complexity of the LINEAR method consists of two components with asymptotic complexity of $\mathcal{O}(KN\rm{log}_2N)$ and $\mathcal{O}(K)$, respectively. The JSPA-WF solution exhibits approximately similar computational complexity to our proposed solution. They differ only in the number of times water-filling is performed. The JSPA-WF solution requires to perform water-filling $N-K$ times while $2N_{\rm{iter}}$ water-filling operations are performed in our proposed solution depending on the adherence quality we desire. The complexity of the ROOT-FIND method  consists of three elements, namely (i) subcarrier allocation  with complexity of $\mathcal{O}(KN)$, (ii) $K$ water-filling operations, and (iii) complexity due to solving iterative root-find method with complexity $\mathcal{O}(nK)$, where $n$ is typically around 10. Therefore, the complexity of our proposed solution is not as high as that of ROOT-FIND, and comparable to LINEAR and JSPA-WF solutions, but provides higher throughput with the best adherence quality.
 
 \section{Conclusions}
\label{section:chapter3-8}
In this letter, we considered an OFDMA system and proposed a near-optimal resource allocation solution that optimizes the system throughput under users' proportional data rate fairness requirements. The proposed solution optimizes the system throughput while strictly maintaining the proportional data rate fairness among the users. Though it is observed that the performance gain of the proposed algorithm is indeed marginal compared to the existing algorithms in terms of achieved throughput, the computational complexity is much lower. Furthermore, our proposed solution performs equally well in both the low-SNR and high-SNR regions. The simulation results also reveal that very strong adherence to the desired proportionality constraints or levels of service is achieved without compromising the system throughput loss. In the present study, the idea is to support mobile users with different levels of service. One could extend this to include machine-to-machine communication scenarios, which would generally have low data rate requirements while operating with lower power level. Furthermore, as a future work, it would be appealing to extend and exploit the ideas in this work to a multi-cell scenario where base station coordination is allowed.

\end{document}